\documentclass[12pt,preprint]{aastex}

\shorttitle{Spin Periods and Magnetic moments in mCVs}
\shortauthors{Norton, Wynn \& Somerscales}

\def\porb{P_{\rm orb}}
\def\pspin{P_{\rm spin}}

\def\rco{R_{\rm co}}
\def\rcirc{R_{\rm circ}}
\def\rin{R_{\rm in}}
\def\ra{R_a}
\def\tdyn{t_{\rm dyn}}
\def\tvisc{t_{\rm visc}}
\def\tmag{t_{\rm mag}}

\begin{document}

%-----------less/greater than approx eq to--------------
\newcommand{\lta}{{\small\raisebox{-0.6ex}{$\,\stackrel
{\raisebox{-.2ex}{$\textstyle <$}}{\sim}\,$}}}
\newcommand{\gta}{{\small\raisebox{-0.6ex}{$\,\stackrel
{\raisebox{-.2ex}{$\textstyle >$}}{\sim}\,$}}}   
%-------------------------------------------------------

\title{The Spin Periods and Magnetic Moments of White Dwarfs in Magnetic 
Cataclysmic Variables}

\author{A.J. Norton}
\affil{Department of Physics and Astronomy, The Open University, Walton Hall,
Milton Keynes MK7 6AA, U.K.}
\email{A.J.Norton@open.ac.uk}

\author{G.A. Wynn}
\affil{Astronomy Group, University of Leicester, Leicester LE1 7RH, U.K.}
\email{gaw@astro.le.ac.uk}

\and

\author{R.V. Somerscales}
\affil{formerly Department of Physics and Astronomy, The Open University, 
Walton Hall, Milton Keynes MK7 6AA, U.K.}

\begin{abstract} 

We have used a model of magnetic accretion to investigate the rotational
equilibria of magnetic cataclysmic variables (mCVs). The results of our
numerical simulations demonstrate that there is a range of parameter space in
the $\pspin / \porb$ versus $\mu_1$ plane at which rotational equilibrium
occurs. This has allowed us to calculate the theoretical histogram describing
the distribution of magnetic CVs as a function of $\pspin / \porb$. We
show that this agrees with the observed distribution assuming
that the number of systems as a function of white dwarf magnetic moment is
distributed approximately according to $N(\mu_1) {\rm d}\mu_1 \propto
\mu_1^{-1} {\rm d}\mu_1$. The rotational equilibria also allow us to infer
approximate values for the magnetic moments of all known intermediate polars. 
We predict that intermediate polars with $\mu_1 \gta 5 \times 10^{33}$~G~cm$^3$
and $\porb > 3$~h will evolve into polars, whilst those with $\mu_1 \lta 5
\times 10^{33}$~G~cm$^3$ and $\porb > 3$~h will either evolve into low field
strength polars which are (presumably) unobservable, and possibly EUV emitters,
or, if their fields are buried by high accretion rates, evolve into
conventional polars once their magnetic fields re-surface when the mass
accretion rate reduces. We speculate that EX Hya-like systems may have low
magnetic field strength secondaries and so avoid synchronisation.  Finally we
note that the equilibria we have investigated correspond to a variety of
different types of accretion flow, including disc-like accretion at small
$\pspin / \porb$ values, stream-like accretion at intermediate $\pspin / \porb$
values, and accretion fed from a ring at the outer edge of the white dwarf
Roche lobe at higher $\pspin / \porb$ values.  

\end{abstract}


\section{Introduction}
\label{sec:intro}

The magnetic cataclysmic variable stars (mCVs) provide a unique insight into
the process of accretion in extreme astrophysical environments. Accretion
within these binary stars takes place in the presence of intense magnetic
fields and high-energy radiation fields. The mCVs are interacting binary stars
in which a magnetic white dwarf (WD) accretes mass from a late-type companion
star via Roche lobe overflow.  The white dwarfs have very large magnetic
moments ($\mu_1 \sim 10^{32} - 10^{35}$~G~cm$^3$) and magnetic stresses have a
pervasive influence on the accretion dynamics. The mCVs fall into two distinct
classes: the AM~Herculis stars (or polars) and the intermediate polars (IPs or
DQ~Herculis stars) (a comprehensive review of mCVs may be found in Warner
1995).  The rotational periods of the WDs ($\pspin$) in the AM~Her stars are
observed to be closely locked to the orbital period ($\porb$), whereas the WDs
within the intermediate polars rotate more rapidly than the orbital period. The
period-locking in the AM~Her stars, which contain the most strongly magnetic
WDs, is thought to come about because the interaction between the magnetic
fields of the two stars in the binary is able to overcome the spin-up torque of
the accreting matter (see e.g. King, Frank and Whitehurst 1991). This suggests
that magnetic effects are important throughout these objects. Indeed,
observations of the AM~Her stars highlight the unique form of the accretion
flow within these binaries, revealing that the gas flow is confined to a stream
flowing along magnetic field-lines on to the poles of the WD. The intermediate
polars represent the various systems which fill the phase space between the
strongly magnetic AM~Her stars and the non-magnetic CVs. The accretion flows
within IPs are known to take on a wide variety of forms, from magnetized
accretion streams to extended accretion discs. This variety has constantly
perplexed efforts to understand these objects.

At some point the accretion flow in all mCVs becomes magnetized and forced into
field-aligned flow.  This magnetized flow reaches the magnetic poles of the
primary star at highly ballistic speeds and passes through a strong shock
before accreting on to the WD.  The hot, post-shock gas is a source of intense
X-ray emission, which is modulated on $\pspin$. Thus, mCVs allow direct
observation of the spin rates of the WDs and provide an unparalleled view of
the angular momentum distribution within the binaries.  Figure 1 shows where
the polars and intermediate polars lie in the $\pspin$ vs. $\porb$ plane;
periods of the intermediate polars are also listed in Table 1.  The current
census included in the figure contains 67 synchronous polars plus 5
asynchronous polars in which $\pspin$ and $\porb$ differ by around 2\% or less.
These latter systems are assumed to have been disturbed from synchronism by
recent nova explosions. Also shown are 39 IPs with a range of properties. Four
systems may best be described as nearly synchronous intermediate polars since
they have $0.9 \gta \pspin / \porb \gta 0.7$; a further six systems lie within
or below the so-called period gap and have $0.7 \gta \pspin / \porb \gta 0.07$;
whilst another six systems with $\pspin / \porb \lta 0.01$ harbour extremely
rapid rotators, including the propeller system AE~Aqr and the very long
outburst interval SU~UMa star, WZ~Sge.  However, the majority of intermediate
polars (currently 23 systems) populate the region defined by $0.25 \gta \pspin
/ \porb \gta 0.01$ and $\porb > 3$~hr.  In this paper we use this period
distribution to investigate the distribution of magnetic moments of the WDs in
mCVs, and the variety of accretion flows that correspond to these equilibrium
spin states.

\section{Magnetic Accretion Flows}
\label{sec:magacc}

The accretion flow between the two stars in a mCV is ionized by collisions as
well as by X-ray irradiation from the strong shock close to the WD surface.
Consequently, the accretion flow is highly conducting and its motion relative
to the magnetic field of the WD will cause magnetic stresses to alter its
dynamics.  The equation of motion of the gas in the binary system can be
written as
\begin{eqnarray}
\label{eqmotion}
\frac{\partial{\bf v}}{\partial t} 
+ ({\rm {\bf v}.\nabla) {\bf v}}  & = & 
- \nabla \Phi_R - 2{\bf \Omega} \wedge {\rm {\bf v}}
-\frac{1}{\rho} \nabla P   \nonumber \\ 
 & & -\frac{1}{\rho} \nabla  \biggl ( \frac{B^2}{8\pi} \biggr ) 
+ \frac{1}{\rho R_c} \biggl ( \frac{B^2}{4\pi} \biggr ) {\rm \hat{\bf n}} 
\end{eqnarray}
where ${\rm \Omega}$ is the orbital angular velocity, $B$ is the local
magnetic field strength, $R_c$ is the local radius of curvature of the
magnetic lines of force and $\Phi_R$ is known as the Roche potential and
includes the effect of both gravitation and centrifugal force.  The term
$-2{\bf \Omega} \wedge {\rm {\bf v}}$ is the Coriolis force per unit mass. The
magnetic field exerts an isotropic pressure $B^2/8\pi$, and carries a tension
$B^2/4\pi$ along the magnetic lines of force.  These magnetic stresses act as
a barrier to the accretion flow, and impede the motion of plasma across field
lines.  The details of the flow of plasma through the magnetosphere are
complex and non-linear. Its motion is dynamic and inherently three dimensional,
unlike the flow in non-magnetic systems which takes the form a planar disc.
The shear velocities between the plasma and the magnetic field are highly
supersonic and in some cases can be relativistic. These features, together
with the plethora of magnetohydrodynamic instabilities which affect the flow,
have made detailed modelling of the accretion process exceedingly difficult. 

In general terms we can characterize accretion flow in the presence of a
magnetic field by three timescales: $t_{\rm dyn} \sim (r^3/GM_1)^{1/2}$ is the
dynamical timescale of the gas orbit, $t_{\rm mag}$ is the timescale on which
magnetic stresses are able to alter the orbital momentum of the accretion flow,
and $t_{visc}$ is the timescale on which viscous stresses are able to
redistribute angular momentum within the gas flow. We can gain a rough estimate
of the extent of the region within which the magnetic field is expected to play
an important part in determining the dynamics of the accretion flow (i.e. where
$\tmag \lta \tdyn$) by comparing the ram pressure of accreting material ($\rho
v^2$) and the local magnetic pressure ($B^2/8\pi$). These two quantities are
defined to be equal at the Alfv\'{e}n radius ($R_a$). Assuming a spherically
symmetric accretion flow without significant pressure support we can deduce the
spherical Alfv\'{e}n radius (e.g. Frank, King \& Raine 2002) 
\begin{equation}
R_a \sim 5.5 \times 10^8 \biggl ( \frac{M_1}{M_\odot} \biggr )^{1/7} R_9^{-2/7}
L_{33}^{-2/7} \mu_{30}^{4/7} \;\; {\rm cm} \label{eqalfven} 
\end{equation}
where $R_9$ is the primary radius (in units of $10^9$ cm), $L_{33}$ is the 
system luminosity (in units of $10^{33}$ erg s$^{-1}$), and $\mu_{30}$ is the 
primary dipole moment (in units of $10^{30}$ G cm$^3$). Typically we find that 
$R_a$ is greater than the system separation ($a$) in the  AM~Her stars, and 
magnetic effects are expected to dominate throughout the binary, as observed.  

In the non-magnetic CVs the accretion process is unimpeded by the presence of a
magnetic field. The accretion flow leaves the secondary star in the form of a
thin stream from its surface (see Warner 1995 for a review of mass transfer via
Roche lobe overflow) and adopts a circular, Keplerian orbit at a radius
$\rcirc$ from the WD. An accretion disc forms from the viscous evolution of the
stream from $\rcirc$ in a time $\tvisc$.  

In the case of a mCV we must consider the effect of the
magnetic field on the disc formation process. If $R_a > \rcirc$ then we have
$\tmag(\rcirc) < \tvisc(\rcirc)$, and magnetic stresses will quickly (in a
time $< \tvisc$) dissipate the stream angular momentum and no accretion disc
will form.  It is possible to estimate $\rcirc$ from the conservation of
angular momentum: $2 \pi b^2 / \porb \sim (G M_1 \rcirc)^{1/2}$, where $b$ is
the distance from the WD to the $L_1$ point and $M_1$ is the mass of the WD.
This results in the relation
\begin{equation}
\frac{\rcirc}{a} \simeq (1 + q) \biggl ( \frac{b}{a} \biggr )^4
\label{rcirc}
\end{equation}
which is a function of the mass ratio $q = M_2/M_1$ only, here $M_2$ is the
mass of the donor star. Typically we find $\rcirc \lta a/10$ in mCVs. As noted
above, $\ra \gg \rcirc$ is satisfied in the AM~Her stars and no accretion disc
can form. In this case we have the hierarchy of timescales $\tmag \lta \tdyn
\ll \tvisc$ at $L_1$.  In the case of the pulsing X-ray binaries, in which a
magnetic neutron star is surrounded by an extensive accretion disc, one finds
that $R_a \ll \rcirc$ and hence $\tmag(\rcirc) \gg\tvisc(\rcirc)$, confirming
that accretion disc formation is unaffected by the magnetic field.  However,
the inward  spread of the accretion disc will eventually be halted by the
magnetic field at some radius $\rin$. The usual definition of $\rin$ is the
point at which the magnetic field removes angular momentum from the disc at a
greater rate than viscous stresses (e.g. Bath, Evans \& Pringle, 1974) 
\begin{equation}
-B_\phi B_z R^2 |_{R_{\rm in}} = \dot{M} \frac{\rm d}{\rm dr} (\Omega_{\rm K}
R^2) |_{R_{\rm in}} \label{rindef} 
\end{equation}
where $B_\phi, B_z$ are the toroidal and poloidal field components
respectively, and $\Omega_{\rm K}$ is the Keplerian angular velocity.  The
value of $\rin$ has an important effect on the equilibrium value of $\pspin$,
which in turn has an important effect on the dynamics of the accretion flow. 
In the case of the asynchronous mCVs we find that in many cases $\ra \gta
\rcirc$, which greatly complicates the accretion dynamics in these systems. A
number of models have been put forward in an attempt to explain these systems. 
The `standard' model of the accretion flow within the intermediate polars
assumes them to be the WD analogue of the pulsing X-ray binaries: accreting via
an extended accretion disc which is disrupted inside $\rcirc$. However it has
become clear (e.g.  Hellier, 1996) that the intermediate polars represent a
variety of different magnetic phenomena. This realization has been prompted by
observational techniques such as Doppler tomography, along with new methods for
the theoretical treatment of the gas flow within the intermediate polars.  

\section{Spin Equilibria}
\label{sec:spin}

Figure 1 shows that the AM~Her systems broadly follow the synchronization line
along which $\pspin = \porb$. The asynchronous systems occupy a wide range of
parameter space between $10^{-3} \lta \pspin / \porb \lta 1$. This fact alone
hints at the variety of magnetic flows which are present within these systems.

The spin periods of IPs allow us a unique insight into the angular momentum
balance within the binary systems, and are a crucial diagnostic tool when
comparing theoretical models. The spin rate of a magnetic WD accreting via a
disc reaches an equilibrium when the rate at which angular momentum is accreted
by the white dwarf is balanced by the braking effect of the magnetic torque on
the disc close to $\rin$. Because of the complex nature of the 
disc-magnetosphere interaction, most models assume that the system is
axisymmetric and in steady state. Many of these models find that the point at
which the accretion disc is disrupted ($\rin$) is very close to the WD
co-rotation radius ($\rco$) (e.g. King \& Lasota 1991). The radius $\rco$ is
defined to be the radius at which the magnetic field rotates at the same rate
as the local Keplerian frequency, and hence $\rco = (GM_1
\pspin^2/4\pi^2)^{1/3}$.  The relation $\rin \sim \rco$ implies that a small
magnetosphere (low $\mu_1$) results in fast equilibrium rotation and vice
versa. Specifically, for a magnetic system with a truncated accretion disc we
expect $\rin \sim \rco \; \ll \; \rcirc$, which translates into the relation
$\pspin/\porb \ll 0.1$.  This is certainly satisfied by the pulsing X-ray
binaries where $\pspin/\porb \lta 10^{-6}$. Figure 1 clearly shows that some of
the asynchronous mCVs also follow this relation. Notably, in GK~Per ($\pspin =
381$ s, $\porb = 48$ hr) $\porb$ is so long that $\ra \ll \rcirc$. The other
rapidly rotating WDs in the systems close to or below the $\pspin = 0.01 \porb$
line presumably have rather low magnetic fields. It would seem then that some
mCVs do indeed conform to the standard model and resemble the pulsing X-ray
binaries in accreting via an extended Keplerian disc, which is disrupted close
to the WD surface. However it is also clear from the figure that this is not
true for all of the systems, as the distribution of spin periods extends out
beyond $\pspin / \porb \gta 0.1$. These systems are unlikely to  possess
accretion discs, since $\pspin/\porb \gta 0.1$ implies  $\rin \sim$ $\rco \gta
\rcirc$. King (1993) and Wynn \& King (1995) examined the regime in which
$\tmag(\rcirc) \lta \tdyn(\rcirc)$ utilizing the model detailed in section 4
below. Numerical calculations show that the WD attains a spin equilibrium
determined approximately by the condition $\rco \sim \rcirc$ corresponding to
$\pspin/\porb \sim 0.1$.  The exact value of the equilibrium $\pspin/\porb$ is
dependent on $q$. In an extension of this analysis King \& Wynn (1999)
discovered a continuum of spin equilibria which explained the long spin period
of EX~Hydra ($\pspin = 67$~min, $\porb = 98$~min; see section 5 below), and
linked all of the mCVs in the context of a single model for the first time. The
fundamental property of these magnetic accretion flows is the non-Keplerian
nature of the velocity field.  

The above arguments implicitly assume that the WDs within the asynchronous mCVs
are in spin equilibrium. This is expected to be the case as the spin-up/down
timescale ($\lta 10^7$~y) of the white dwarf is short when compared to the
lifetime of the binary ($\gta 10^9$~y), (see e.g. Wynn \& King 1995). We also
note that in cases where $\dot{P}_{\rm spin}$ is measured for intermediate
polars, several are spinning up whilst a similar number are spinning down, and
FO Aqr has been seen to do both (e.g. Patterson et al 1998). Hence the
assumption that many intermediate polars {\em are} close to their equilibrium
spin periods is a good one with only small deviations apparent on short
timescales. However, this is certainly not the case in all of the systems: the
very short spin period of AE~Aqr ($\pspin = 33$ s, $\porb = 9.88$ hr) does not
imply the presence of a Keplerian accretion disc as the WD is currently a long
way from spin equilibrium (Wynn, King \& Horne 1997).  

\section{The Magnetic Model}

The equation of motion for gas flow within mCVs (e.g. Equation \ref{eqmotion})
includes magnetic pressure and tension terms. King (1993) and Wynn \& King
(1995) constructed a model of accretion flows in the mCVs by assuming that
material moving through the magnetosphere interacts with the local magnetic
field via a shear-velocity dependent acceleration. This is analogous to the
assumption that the magnetic stresses are dominated by the tension term in
Equation \ref{eqmotion} giving a magnetic acceleration
\begin{equation}
{\bf a}_{\rm mag} \simeq \frac{1}{R_c \rho} \biggl (
\frac{B^2}{4\pi} \biggr ) {\rm \hat{\bf n}}
\end{equation}
Where $R_c$ is local radius of curvature of the magnetic lines of force, $B$ is
the local magnetic field strength, and $\rho$ is the local gas density. Because
the magnetic tension stresses arise from currents induced by the rotational
shear between the plasma flow and the external WD magnetosphere, we can write
this acceleration in terms of the velocity shear in the general form
\begin{equation}
{\bf a}_{\rm mag} \simeq -k(r,t,\rho) [\textbf{v}-\textbf{v}_{f}]_{\bot}
\end{equation} 
where $\textbf{v}$ and  $\rm \textbf{v}_{f}$ are the velocities of the material
and field lines, and the suffix $\bot$ refers to the velocity components
perpendicular to the field lines. The coefficient $k$ contains the details of
the plasma-magnetic field interaction, and will be a function of position, time
and gas density in general. This can be found, at least in approximate form,
for the limiting cases of diamagnetic and magnetized accretion flows. Recently,
for example, Matthews, Speith \& Wynn (2004) used a similar treatment to study
the effect of a central magnetic field on the structure of a magnetized
accretion disc in the FU~Orionis objects.  Some of the advantages of this
approach are that it simplifies the complex MHD, is easy to add in to
hydrodynamic codes, and allows investigation of the spin evolution of WDs in
mCVs. The main drawbacks are that some physics is inevitably lost and the $k$
parameter is model dependent. King (1993) and Wynn \& King (1995) approximated
the plasma flow in mCVs as inhomogeneous and diamagnetic, interacting with the
field via a surface drag force characterized by the magnetic timescale
\begin{equation}
\label{tmag} t_{\rm mag}\sim c_{\rm A}\rho_{\rm b}l_{\rm b}B^{-2}
\frac{|\textbf{v}_{\bot}|}{|\textbf{v}-\textbf{v}_{f}|_{\bot}} 
\end{equation}
(cf.\ Equation 10 in  Matthews, Speith \& Wynn 2004).
The $k$ parameter in this case is given by 
\begin{equation}
k \sim \frac{B^2}{c_{\rm A} \rho_{\rm b} l_{\rm b}} \sim t_{\rm mag,slow}^{-1}
\end{equation}
where $c_{\rm A}$ is the Alfv\'{e}n speed in the medium surrounding the plasma,
$\rho_{\rm b}$ is the plasma density and $l_{\rm b}$ is the typical
length-scale over which field-lines are distorted. Here $t_{\rm mag,slow}$ is
the limiting value of $t_{\rm mag}$ in the case of a slowly rotating WD
($\textbf{v}_{f} \rightarrow 0$).  The plasma density and length-scale may be
approximated by
\begin{equation}
\rho_{\rm b} \sim \frac{\dot{M} 4 \pi^2}{P^2_{\rm orb} c^3_{\rm s}}
\end{equation}
and
\begin{equation}
l_{\rm b} \sim \frac{P_{\rm orb} c_{\rm s}}{2 \pi}
\end{equation}
which are the accretion stream density and width in the vicinity of the L1 point
respectively, and $c_{\rm s}$ is the sound speed. Plasma will exchange orbital
energy and angular momentum with the field on the magnetic timescale, which is
dependent on $\pspin$ since $|\textbf{v}_{f}| \sim 2\pi R / \pspin$.  Material
at radii greater than $\rco$ will experience a net gain of angular momentum
and be ejected from the binary or captured by the secondary star. On the other
hand, material inside $\rco$ will lose angular momentum and be accreted by the
WD. An equilibrium will result when these angular momentum flows balance. 
Hence the spin evolution of the WD in mCVs can be followed in detail. 

The results presented below were obtained from spin evolution calculations
performed using a 3D particle hydrodynamics code using an implementation of the
magnetic model described above. The accretion flow calculations were performed
in the full binary potential, and included a simple treatment of the gas
viscosity. This treatment used a constant viscosity parameter, calibrated to
give similar results to a Shakura \& Sunyaev parameter of $\alpha \sim 0.01$
for accretion disc formation simulations (i.e.\ very low $k$). This value was
chosen as appropriate for cold state gas flow (i.e.\ $\lta 10^4$ K as in
quiescent dwarf novae discs) which is the expected state of the accretion flow
away from the WD surface in mCVs. The reader is referred to Wynn \& King (1995,
and references therein) for further details on this code.  

\section{Results}

The magnetic model has been used to explore a wide range of IPs parameterized
by their orbital period ($P_{\rm orb} = 80$ min to 10 hours) and magnetic field
strength ($k = (10^2 - 10^7) (2\pi / \porb) {\rm s}^{-1}$), using $M_1 =
0.7$M$_{\odot}$ and $M_2 = 0.35$M$_{\odot}$ (i.e. $q = 0.5$). The mass transfer
rate and WD moment of inertia were chosen to facilitate progress towards the
equilibrium spin state.  All models assume a dipolar structure for the WD
magnetic field.  

The equilibrium spin periods we have determined are shown by the small dots in
Figure 2. Here the $k$ values have been converted into the corresponding value
of the white dwarf surface magnetic moment $\mu_1$ using Equations (8), (9) and
(10) above.  Lines on Figure 2 connect the dots for each orbital period. The
line for $P_{\rm orb} =$ 80~m is similar to that obtained by King \& Wynn
(1999) for EX Hya, although here calculated for a different mass ratio. It
should be noted that there is some latitude in applying the conversion from $k$
to $\mu_1$, so the envelope of equilibrium periods may shift laterally in the
plane.  Nonetheless it is encouraging that the predicted equilibrium spin
periods lie in the range at which many mCVs are seen.  It should also be noted
that the estimated magnetic field strengths for the intermediate polars that
display polarized emission (e.g. PQ Gem, BG CMi, V2400 Oph and LS Peg) lie
within the region covered by the results.

The primary source of uncertainty associated with the equilibrium spin periods
in Figure 2 is $q$. Broadly speaking, the overall shape of the envelope
encompassed by the curves on Figure 2 is independent of $q$. However, the
curves shift verically in the figure with changes in the mass ratio. The
reason for this can be seen from the dependence
\begin{equation}
\frac{\pspin}{\porb} \propto \biggl ( \frac{b}{a} \biggr )^{3/2} \propto (0.500 - 0.227\log q)^{3/2}
\label{eqqrel}
\end{equation}
which can be deduced from Equation 3 (see also Equation 13 in King \& Wynn
1999). This relation predicts a variation in the predicted equilibrium spin
values of $\lta 40$\% when considering the extremes of the expected mass ratio
range in CVs ($0.1 \lta q \lta 0.9$). For mass ratios between the more 
conservative limits of $0.3 \lta q \lta 0.7$, the variation in the predicted
equilibrium spin period is only of order 10\%. These results are borne out 
to a high degree of accuracy by numerical calculation.  

If we now assume that all mCVs are at their equilibrium spin period we can
place the intermediate polars on Figure 2 by tracing across at the appropriate
$P_{\rm spin}/P_{\rm orb}$ value to intersect the equilibrium line for the
appropriate orbital period. In this way, most of the squares have been placed
on Figure 2. By doing this we `predict' a value for the surface magnetic moment
$\mu_1$ of the white dwarf in each system; and these values are listed in Table
1. Given the latitude in converting from $k$ to $\mu_1$, and also the
dependence of the location of these curves on $q$, we emphasize that these
$\mu_1$ values should only be relied upon as an order of magnitude indication
of the value in each case.

We note here that since our results were all obtained with $q=0.5$, they
preclude fitting the exact spin states of individual systems in detail. For
example, the spin period of EX~Hya lies above all of the equilibrium curves we
present. However, King \& Wynn (1999) were able to fit the spin of EX~Hya with
an identical model, but using the system parameters specific to that object. 
The principal difference between the parameters they adopted and the ones we
used here is the difference in the mass ratio of the system. The consequent
difference in equilibrium spin period can be predicted from the discussion
presented above. The mass ratio of EX Hya is $q = 0.17^{+0.08}_{-0.07}$ 
(Beuermann et al 2003), so Equation (\ref{eqqrel}) yields the relation 
\begin{equation}
\biggl ( \frac{\pspin}{\porb} \biggr )_{q = 0.17} \simeq 1.30^{+0.15}_{-0.11}
\biggl ( \frac{\pspin}{\porb} \biggr )_{q = 0.5} 
\end{equation} 
Since $(\pspin / \porb)_{q = 0.5} \sim 0.54$ from our results, we 
predict $(\pspin/\porb)_{q=0.17} \sim 0.70^{+0.08}_{-0.06}$ for EX~Hya, which 
is indeed consistent with what is observed and the result of King and Wynn 
(1999). 

For the four nearly synchronous intermediate polars that have $\pspin / \porb
\gta 0.7$  we have simply placed them on Figure 2 above the mid-point of the
plateau part of the relevant curves at the appropriate $\pspin / \porb$ value. 
If these systems are in equilibrium, a similar argument to that used for EX Hya
shows that they would have extremely low mass ratios ($q \lta 0.15$). For this
reason, we prefer to identify these systems as being out of equilibrium and 
heading towards synchronism.

In the envelope of allowed parameter space shown on Figure 2, we note that the
region with $P_{\rm spin}/P_{\rm orb} \lta 0.1$ corresponds to $R_{\rm mag}
\sim R_{\rm co} \lta R_{\rm circ}$ and so indicates regions where truncated
accretion discs are likely to form. In fact, for $\pspin / \porb \lta 0.1$ the
magnetic model shows that equilibrium spin periods do indeed produce accretion
flows that are disc-like in nature, whereas for $0.1 \lta \pspin / \porb \lta
0.5$ stream-like accretion is the preferred mode. Figure 3 shows examples of
some of these accretion flows, corresponding to different regions of the
equilibrium parameter space. At even higher period ratios of $\pspin / \porb
\gta 0.5$, we find solutions in which the accretion is fed from a ring-like
structure at the outer edge of the WD Roche lobe (Figure 3).  This mode of
accretion flow is apparent in the upper plateau region of the curves shown in
Figure 2 and corresponds to the situation $R_{\rm mag} \sim R_{\rm co} \sim b$.
We note that this is an additional solution to that found by King \& Wynn
(1999) for EX Hya. In their solution, systems undergo alternate phases of
accretion and ejection to maintain equilibrium whilst fed from a stream. In the
solution we have found here, angular momentum from the white dwarf is passed
back to the accreting material and some of this material is lost from the outer
edge of the ring to maintain equilibrium. This ring-fed solution may be
preferred over the stream-fed accretion/ejection solution when the angle
between the magnetic dipole axis and the spin axis is small. In fact, the
ring-fed solution may be more consistent with the observational data on EX Hya
than King \& Wynn's previous solution, since some observations do indeed show
evidence for material circling the white dwarf in that system.

\section{Synchronisation}

We now use the results obtained above to investigate the polar synchronisation
condition. We may assume that systems become synchronized once the magnetic
locking torque is equal to the accretion torque. Hence
\begin{equation}
\label{synch}
\frac {\mu_1 \mu_2}{a^3} = \dot{M} ( G M_1 R_{\rm mag})^{1/2}
\end{equation}
where the surface magnetic moment on the secondary may be approximated by
\begin{equation}
\label{mu2}
\mu_2 = 2.8 \times 10^{33} P_{\rm orb}^{9/4}~{\rm G~cm}^3
\end{equation}
from Warner (1996) with $P_{\rm orb}$ in hours. The magnetospheric radius
is approximated by the co-rotation radius, 
\begin{equation}
R_{\rm mag} \sim R_{\rm co} = \left( \frac{G M_1 P_{\rm spin}^2} {4\pi^2} 
\right)^{1/3}
\end{equation}
and the mass accretion rate $\dot{M}$ is set to the secular value given by 
\begin{equation}
\dot{M} = 2.0 \times 10^{-11} 
\times P_{\rm orb}^{3.7}~{\rm M}_{\odot}~{\rm yr}^{-1}
\end{equation}
for $P_{\rm orb} > 2.7$~h (McDermott \& Taam 1989) and 
\begin{equation}
\dot{M} = 2.4 \times 10^{15} \times \frac{M_1^{2/3} P_{\rm orb}^{-1/6}}
{\left(1 - \frac{15q}{19}\right) \left(1+q\right)^{1/3} }~{\rm g~s}^{-1}
\end{equation}
for angular momentum loss by gravitational radiation (Warner 1995) which
we assume to be dominant at $P_{\rm orb} < 2.7$~h. 

Solutions to Equation \ref{synch} are plotted as diagonal lines on Figure 4.
Each line connects points at which synchronisation will occur for a given
orbital period. The lines shown are for orbital periods of 3~h, 4~h, ... 10~h,
as those for shorter orbital periods lie to the upper left of the region
plotted. Where these diagonal lines intersect the equilibrium spin period lines
for their respective orbital periods, gives the points at which systems will
synchronize. The locus of these points is shown by the thick line which simply
connects the various intersection points. Hence, systems lying below the thick
line will not be prone to synchronize, whilst those lying above will tend to do
so.

\section{Predictions}

From Figures 2 and 4 it can be seen that the intermediate polars above the
period gap virtually all lie {\em below} the synchronisation line. The two
systems just above the line are RR Cha and RXJ0944.5+0357 with $P_{\rm
spin}/P_{\rm orb}$ of 0.161 and 0.168 respectively. We suggest these two are
probably normal intermediate polars and within the errors of our calculation
are consistent with sitting below the synchronisation line. Lying well above
the synchronisation line are the two nearly synchronous intermediate polars
V697 Sco and HS0922+1333 with $P_{\rm spin}/P_{\rm orb}$ of 0.737 and 0.882
respectively. As noted above, we suggest that these two systems are on their
way to synchronism. A similar fate should hold for the rest of the intermediate
polars above the period gap that have relatively high white dwarf magnetic
moments.

At short orbital periods, the picture is rather different. First, we note that
there is a relative dearth of systems here. The number of non-magnetic CVs
above the period gap is comparable to the number below it (Ritter \& Kolb
2003), yet there are at least four times as many intermediate polars above the
gap as below it. We suggest that the original evolutionary arguments for
magnetic CVs were probably correct -- the majority of intermediate polars do
indeed evolve to synchronism and become polars, so explaining why there are so
few intermediate polars at short orbital periods. All 6 of the intermediate
polars that are observed below or within the period gap (`EX Hya-like
systems'), plus the two nearly synchronous intermediate polars RXJ0524+42 and
V381Vel, lie above the synchronisation line and so {\em should} be
synchronized. We suggest the reason for them not being so is simply  that the
secondary stars in the EX Hya-like systems have low magnetic moments (i.e. less
than predicted by Equation \ref{mu2}) and so are unable to come into
synchronism. They therefore represent the low magnetic field tail-end of a
distribution.  

It is also apparent from Figure 2 that virtually all the intermediate polars
have $\mu_1 \lta 5 \times 10^{33}$~G~cm$^3$ whereas virtually all the polars
have $\mu_1 \gta 5 \times 10^{33}$~G~cm$^3$. This discrepancy between the
intermediate polar and polar magnetic fields has been noted in the past and was
a reason for rejecting the hypothesis of polars evolving into intermediate
polars. It also poses the question: where are the synchronous systems with low
white dwarf magnetic moments? We can suggest two possibilities for the answer. 
Firstly, it may be that when the systems with a low white dwarf magnetic moment
reach a synchronous state they become primarily EUV emitters and so are
unobservable.  Secondly there is the possibility raised by Cumming (2002) that
the magnetic fields in intermediate polars are buried by their high accretion
rate and so are not really as low as they appear. In this picture, as
intermediate polars evolve towards shorter orbital periods, mass transfer shuts
off when they reach the period gap, the magnetic field of the white dwarf
re-surfaces, and the systems synchronize before re-appearing below the period
gap as conventional, high field, polars.

\section{The magnetic moment distribution of WDs in mCVs}

The results shown in Figure 4 can also be used to characterize the theoretical
distribution of mCVs as a function of the white dwarf surface magnetic moment.
We assume that the number of systems varies according to
\begin{equation} 
\label{mu1}
N(\mu_1) {\rm d}\mu_1 \propto \mu_1^{-n} {\rm d}\mu_1
\end{equation}
where $n$ is a number to be determined. We may integrate under the 
equilibrium curves shown in Figure 4 to get the predicted number of systems
within a given range of $P_{\rm spin}/P_{\rm orb}$. The number in a given
bin is then
\begin{equation}
N \propto \left[  \frac{\mu_1^{1-n}}{1-n}  \right] ^{\mu_1}
_{\mu_1 + \frac {\Delta \frac{P_{\rm spin}}{P_{\rm orb}}}{G}}
\end{equation}
where $G$ is the gradient of the equilibrium curves in Figure 4 at any
point. The range of spin to orbital period ratios was divided into 
eight logarithmic bins, seven of which were 0.25 wide in log space and
the eighth corresponded to synchronous systems. The integration
was carried out over the magnetic moment range $10^{32}~{\rm G~cm}^3 <
\mu_1 < 10^{35}~{\rm G~cm}^3$ and at the spin to
orbital period ratio (for a given orbital period) indicated by the thick line 
in Figure 4, we assume that systems become synchronized. The predicted number
of systems was averaged over orbital periods of 3, 4, 5, and 6 hours
assuming systems to be distributed evenly in orbital period, and the final 
result was normalized to the number of known mCVs with orbital periods
greater than 3~h. 

The best fit value of the power law index $n$ was found to be at 1.10, with a
reduced chi-squared of $\chi^2_r = 0.44$, as shown in Figure 5, although values
between $\sim$0.95 and 1.27 are also valid ($\chi^2_r < 1$).  The cumulative
histograms comparing the observed number of systems with that predicted by
Equation  \ref{mu1} with $n=1.10$ are shown in Figure 6, and the data are 
listed in Table 2. A power law index of $n \sim 1$ indicates roughly equal 
numbers of systems per decade of magnetic moment.

\section{Testing the applicability of the model}

As we have demonstrated, our magnetic model currently evolves systems to their
equilibrium spin period and allows examination of the resulting accretion flow.
In order to assess the validity of the model, and how accurately it reproduces
accretion flows seen in real IPs, it is necessary to test the observable
consequences of these results against real data. One such test will be to 
investigate which of the model flows can support dwarf nova instabilities,
evidence for which is seen in a few IPs. We also plan to extend the
scope of the model in order to produce synthetic observables which can then
be compared with, or fitted to, the wealth of existing data on individual
systems. One part of this approach will be to produce synthetic X-ray
lightcurves and power spectra, the other will be to produce synthetic optical
Doppler tomograms and trailed spectra.

The model explicitly tracks the position and velocity of each diamagnetic
blob from the L1 point down to the WD surface. It is therefore already possible
to output the mass accretion rate onto a given pole as a function of time, and
this capability was illustrated in our recent discussion of RXJ1914.4+2456 and
RXJ0806.3+1527 as face-on, stream-fed IPs (Norton, Haswell \& Wynn 2004). But
this is {\em not} the same as predicting the X-ray lightcurve. The construction
of synthetic X-ray lightcurves from the position and velocity information as a
function of time for the accreting material will firstly involve modelling the
spectrum and intensity of X-ray emission of an individual blob as it undergoes
a shock in the vicinity of the WD surface. The next step will be to perform a
`ray tracing' operation to track back the X-ray emission along the line of
sight, accounting for occultation of the emission as well as photoelectric
absorption and electron scattering in the intervening blobs. By summing the
emission from all blobs and allowing the system to evolve in time, synthetic
lightcurves (and hence power spectra) can be built up exhibiting variation as a
function of white dwarf spin phase, binary phase, and beat phase, all as a 
function of X-ray energy.

We can also use the accretion flows produced by the model to generate
synthetic maps of optical emission by assigning a dissipation rate to each
particle in the flow.  Since we already know the velocity of each particle,
such emission maps can then be converted into synthetic optical trailed
spectra, phased both with the spin and orbital periods of a given system.
The synthetic trailed spectra can then be convereted into synthetic Doppler
tomograms i.e. maps of the emission in velocity space, for comparison
with observational data.

\section{Conclusions}

We have demonstrated that there is a range of parameter space in the $\pspin / 
\porb$ versus $\mu_1$ plane at which spin equilibrium occurs for
mCVs. At equilibrium, with a mass ratio of $q=0.5$, for $\pspin / \porb \lta
0.1$ accretion is via a truncated accretion disc; for $0.1 \lta \pspin / \porb
\lta 0.5$ stream-fed accretion occurs; whilst for $\pspin / \porb \gta 0.5$
accretion may be fed from a ring at the outer edge of the WD Roche lobe. Varying
the mass ratio shifts the boundaries of these types of behaviour by of order
ten percent, but the overall pattern remains the same.

Using the results of numerical simulations we can infer the surface magnetic
moments of the white dwarfs in intermediate polars to be mostly in the range
$10^{32}~{\rm G~cm}^3 \lta \mu_1 \lta 5 \times 10^{33}~{\rm G~cm}^3$.  Most of
the intermediate polars with orbital periods greater than 3~h lie below the
synchronisation line, whilst most systems with orbital periods less than 2~h
lie above it and should be synchronized. High magnetic moment intermediate
polars at long orbital periods will evolve into polars. Low magnetic moment
intermediate polars at long orbital periods will either: evolve into low field
strength polars which are (presumably) unobservable, and possibly EUV emitters;
or, if their fields are buried by high accretion rates, evolve into
conventional polars once their magnetic fields re-surface when the mass
accretion rate reduces. EX Hya-like systems may represent the tail end of a
distribution such that they have low magnetic field strength secondaries and so
avoid synchronisation.

We have also shown that the distribution of mCVs above the period gap follows
the relationship $N(\mu_1) {\rm d}\mu_1 \propto \mu_1^{-1} {\rm d}\mu_1$,
indicating roughly equal numbers of systems per decade of magnetic moment.  For
comparison, recent data from the Sloan Digital Sky Survey (Schmidt et al 2003),
concerning the magnetic field strength distribution of isolated white dwarfs,
reveals 38 systems with $10^6 < B / {\rm G} < 10^7$, 49 systems with $10^7 < B
/ {\rm G} < 10^8$ and 20 systems with $10^8 < B / {\rm G} < 10^9$.  

Future work will assess the validity of the accretion flows we have simulated
by comparing synthetic observables generated from these flows with the wealth
of existing data on individual systems.

\section{Acknowledgements}
This research has made use of computing facilities in the Open University
Department of Physics and Astronomy which are funded by the OU Research
Development Fund and Capital Equipment Fund, and supported by PPARC
grant PPA/G/O/2000/00037. Theoretical astrophysics at Leicester is supported 
by a PPARC rolling grant.

\clearpage

\begin{deluxetable}{lrccccl}
\tabletypesize{\scriptsize}
\tablewidth{0pt}
\tablecaption{Intermediate polars with known spin and orbital periods}
\tablehead{
\colhead{Star}	& \colhead{X-ray name}	& 
\colhead{$P_{\rm spin}$/s} & \colhead{$P_{\rm orb}$/h} & 
\colhead{$P_{\rm spin}/P_{\rm orb}$} & 
\colhead{Inferred  $\mu_1$/G cm$^3$} & \colhead{Period Ref.}}
\startdata
\multicolumn{2}{l}{\bf Rapid rotators}  & & & & \\
WZ~Sge		& 		       &    28.96    & 1.3606	 &  0.0572	  & $<10^{32}$ & 1\\
AE~Aqr          & 1E2037.5-0102        &    33.0767  & 9.87973   &  0.00093       &  $<10^{32}$ & 2\\
GK~Per          & 3A0327+438           &    351.341  & 47.9233   &  0.00204       &  $<10^{32}$ & 3\\
V533~Her        &                      &    63.633   & 3.53      &  0.00501       &  $<10^{32}$ & 4\\
DQ~Her          &                      &    142      & 4.6469    &  0.00849       &  $<10^{32}$ & 5\\
XY~Ari          & H0253+193            &    206.30   & 6.0648    &  0.00945       &  $4.5\times10^{32}$ & 6\\
\multicolumn{2}{l}{\bf Regular IPs, $P_{\rm orb} > 3$h} & & & & \\
V709~Cas        & RXJ0028.8+5917       &    312.8    & 5.34      &  0.0163        &  $4.2\times10^{32}$ & 7\\
                & 1RXSJ1548-4528       &    693      & 6.72/9.37 &  0.0286/0.0205 &  $1.0/1.4\times10^{33}$ & 8\\
2236+0052       &                      &    403.7    & 3.23      &  0.0347        &  $2.6\times10^{32}$ & 9\\
V405~Aur        & RXJ0558.0+5353       &    545      & 4.15      &  0.0365        &  $4.8\times10^{32}$ & 10\\
YY~Dra          & 3A1148+719           &    529.31   & 3.96898   &  0.0370        &  $4.7\times10^{32}$ & 11\\
PQ~Gem          & 2REJ0751+144         &    833.411  & 5.1927    &  0.0446        &  $8.3\times10^{32}$ & 12\\
V1223~Sgr       & 4U1849-31            &    745.506  & 3.36586   &  0.0615        &  $4.0\times10^{32}$ & 13\\
AO~Psc          & H2252-035            &     805.203 & 3.59105   &  0.0623        &  $5.2\times10^{32}$ & 13\\
HZ~Pup          &                      &    1212.2   & 5.11      &  0.0659        &  $9.1\times10^{32}$ & 14\\
UU~Col          & RXJ0512.2-3241       &    863.5    & 3.45      &  0.0695        &  $6.5\times10^{32}$ & 15\\
                & 1RXSJ062518.2+733433 &    1187.244 & 4.71863   &  0.0699        &  $1.0\times10^{33}$ & 16\\    
FO~Aqr          & H2215-086            &    1254.451 & 4.84944   &  0.0718        &  $1.2\times10^{33}$ & 17\\
V2400~Oph       & RXJ1712.6-2414       &    927      & 3.41      &  0.0755        &  $1.1\times10^{33}$ & 18\\
BG~CMi          & 3A0729+103           &    913.496  & 3.23397   &  0.0785        &  $1.2\times10^{33}$ & 19\\
TX~Col          & H0542-407            &    1911     & 5.718     &  0.0928        &  $4.0\times10^{33}$ & 20\\
                & 1WGA1958.2+3232      &    1466.66  & 4.35      &  0.0937        &  $2.5\times10^{33}$ & 21\\
TV~Col          & 3A0527-329           &    1910     & 5.48641   &  0.0967        &  $4.1\times10^{33}$ & 22\\
AP~Cru          &                      &    1837     & 5.12      &  0.0997        &  $3.7\times10^{33}$ & 23\\
V1062~Tau       & H0459+246            &    3726     & 9.952     &  0.104         &  $1.4\times10^{34}$ & 24\\
LS~Peg          &                      &    1776     & 4.19      &  0.118         &  $3.4\times10^{33}$ & 25\\
RR~Cha          &                      &    1950     & 3.37      &  0.161         &  $3.4\times10^{33}$ & 23\\
                & RXJ0944.5+0357       &    2162     & 3.58      &  0.168         &  $4.0\times10^{33}$ & 9\\
V1425~Aql       &                      &    5188     & 6.14      &  0.235         &  $1.8\times10^{34}$ & 26\\
\multicolumn{2}{l}{\bf EX Hya-like IPs, $P_{\rm orb} < 3$h} & & & & \\
DD~Cir          &                      &    670      & 2.34      &  0.080         &  $1.7\times10^{32}$ & 27\\
HT~Cam          & RXJ0757.0+6306       &    511      & 1.35      &  0.105         &  $2.7\times10^{32}$ & 28\\
V795~Her	&	 	       &    1172     & 2.60      &  0.125         &  $3.6\times10^{32}$ & 29\\
                & RXJ1039.7-0507       &    1444     & 1.574     &  0.255         &  $9.2\times10^{32}$ & 30\\
V1025~Cen       & RXJ1238-38           &    2147     & 1.42      &  0.420         &  $1.8\times10^{33}$ & 31\\
EX~Hya          & 4U1249-28            &    4021.62  & 1.637612  &  0.682         &  $\sim 5\times10^{33}$ & 32\\
\multicolumn{2}{l}{\bf Nearly synchronous IPs} & & & & \\
V697~Sco        &                      &    11916    & 4.49      &  0.737         &  $>1\times10^{35}$ & 33\\
HS0922+1333     &                      &    14636.16 & 4.608     &  0.882         &  $>1\times10^{35}$ & 34\\ 
                & RXJ0524+42           &    8160     & 2.617     &  0.866         &  $>1\times10^{34}$  & 35\\
V381~Vel        &                      &    7320     & 2.233     &  0.910         &  $>1\times10^{34}$  & 34\\
\enddata
\tablerefs{
(1) Welsh et al 2003; (2) Choi, Dotani \& Agrawal 1999; 
(3) Morales-Rueda, Still \& Roche 1996; (4) Thorstensen \& Taylor 2000; 
(5) Zhang et al 1995; (6) Hellier, Mukai \& Beardmore 1997; 
(7) de Martino et al 2001; (8) Haberl, Motch \& Zickgraf 2002; 
(9) Woudt \& Warner 2004; (10) Harlaftis \& Horne 1999; 
(11) Haswell et al 1997; (12) Duck et al 1994; 
(13) Taylor et al 1997; (14) Abbott \& Shafter 1997; 
(15) Burwitz et al 1996; (16) Staude et al 2003;
(17) de Martino et al 1999; (18) Buckley et al 1997;
(19) de Martino et al 1995; (20) Norton et al 1997;
(21) Norton et al 2002; (22) Vrtilek et al 1996;
(23) Woudt \& Warner 2002; (24) Hellier, Beardmore \& Mukai 2002;  
(25) Rodriguez-Gil et al 2001; (26) Retter, Leibowitz \& Kovo-Kariti 1998;
(27) Woudt \& Warner 2003b; (28) Kemp et al 2002; 
(29) Rodriguez-Gil et al 2002; (30) Woudt \& Warner 2003a; 
(31) Hellier, Wynn \& Buckley 2002; (32) Allan, Hellier \& Beardmore 1998; 
(33) Warner \& Woudt 2002; (34) Tovmassian 2003;
(35) Schwarz 2003.}
\end{deluxetable}

\clearpage

\begin{deluxetable}{cccc}
\tablewidth{0pt}
\tablecaption{Histograms comparing the observed number of systems with
that predicted by Equation \ref{mu1} with $n=1.10$. The first two columns
indicate the lower edge of the bins.}
\tablehead{
\colhead{log($P_{\rm spin}/P_{\rm orb}$)} &  
\colhead{$P_{\rm spin}/P_{\rm orb}$} & 
\colhead{$N$(obs)} & \colhead{$N$(model)}}
\startdata
--2.00 & 0.010 & 1  & 0.44\\
--1.75 & 0.018 & 1  & 0.72\\
--1.50 & 0.032 & 4  & 1.34 \\
--1.25 & 0.056 & 12 & 17.77 \\
--1.00 & 0.100 & 4 & 1.74 \\
--0.75 & 0.178 & 1  & 1.26 \\
--0.50 & 0.316 & 0  & 0.32 \\
--0.25 & 0.562 & 2  & 0.00 \\
0.00  & 1.000 & 26 & 27.41 \\
\enddata
\end{deluxetable}

\clearpage

\begin{figure} 
\includegraphics[angle=-90,scale=0.6]{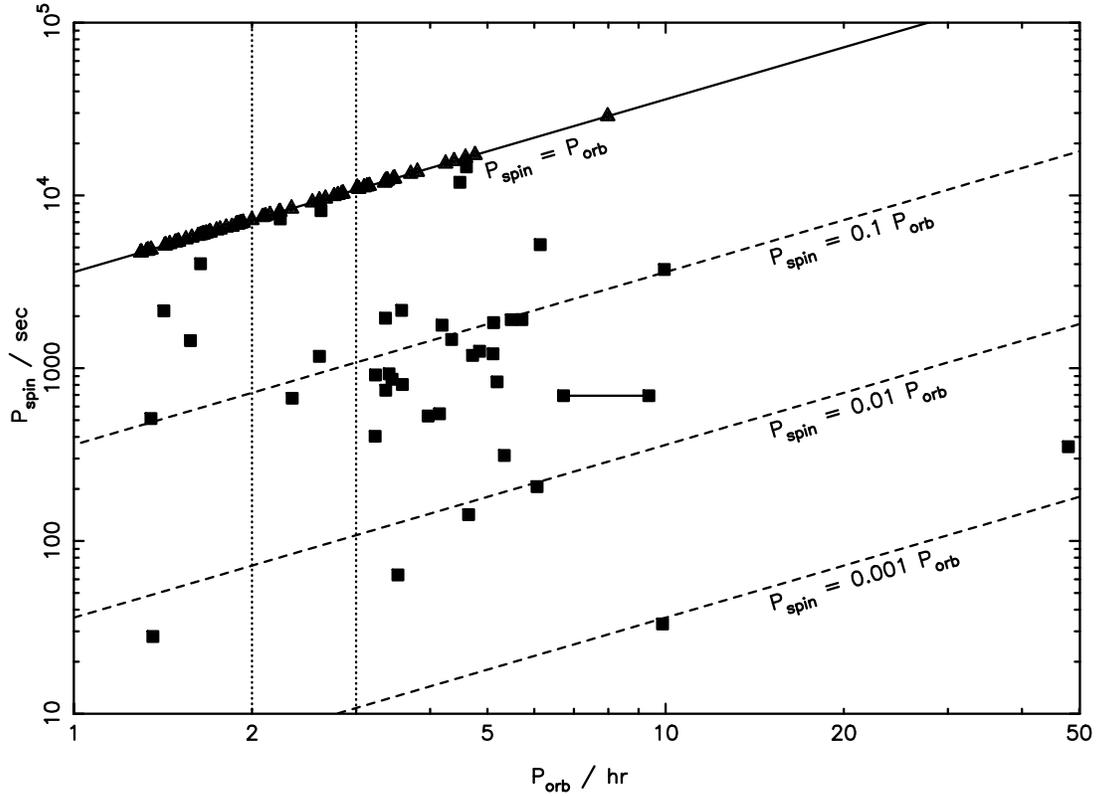}
\caption{The spin and orbital periods of the mCVs. Polars are indicated by 
triangles and intermediate polars by squares. The four `nearly synchronous' IPs
(i.e. $0.9 > \pspin / \porb > 0.7$) are V697 Sco, HS0922+1333, 
RX~J0425+42 and V381 Vel; the six `rapid rotators' shown (i.e. $\pspin / 
\porb < 0.01$) are WZ Sge, AE Aqr, GK Per, V533 Her, DQ Her and XY Ari; the 
six conventional IPs within or below the period gap are DD Cir, HT Cam, 
V795 Her, RX~J1039.7--0507, V1025 Cen and EX Hya. The remaining 23 
conventional IPs above the period gap are V709 Cas, 1RXSJ154814.5-452845, 
2236+0052, V405 Aur, YY Dra, PQ Gem, V1223 Sgr, AO Psc, HZ Pup, UU Col, 
1RXSJ062518.2+733433, FO Aqr, V2400 Oph, BG CMi, TX Col, 1WGA1958.2+3232, 
TV Col, AP Cru, V1062 Tau, LS Peg, RR Cha, RXJ0944.5+0357 and V1425 Aql. 
See also Table 1.}
\end{figure}

\clearpage

\begin{figure}
\includegraphics[angle=-90,scale=0.6]{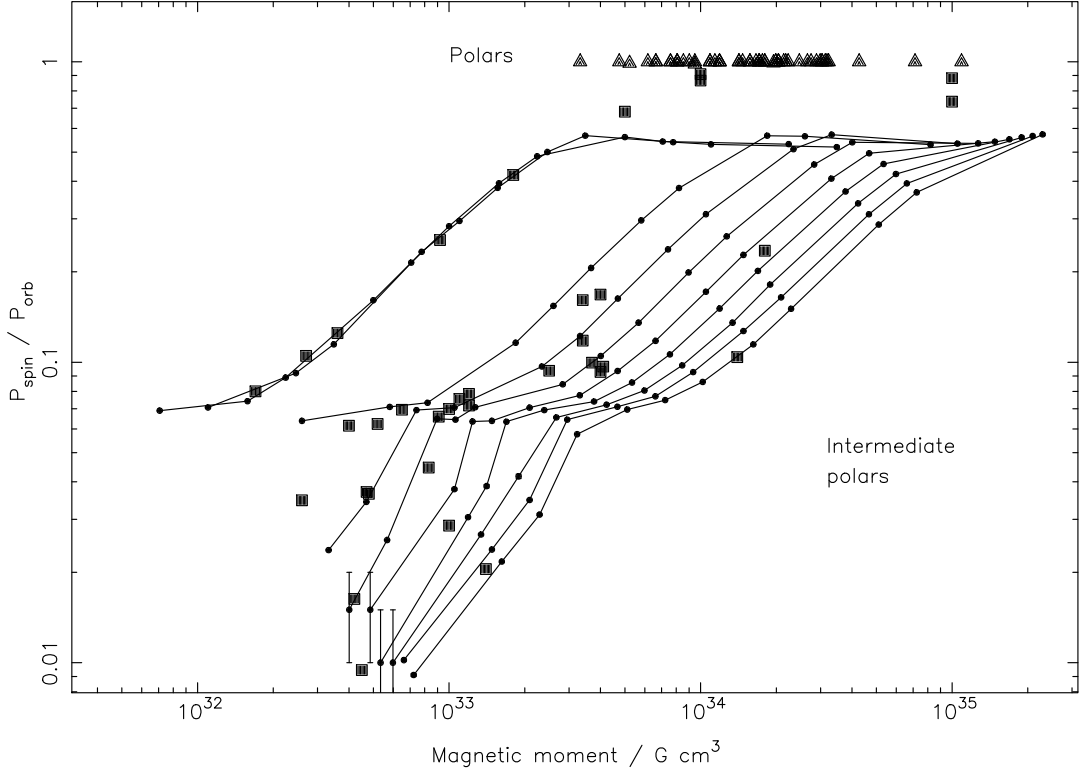}
\caption{The equilibrium spin periods of mCVs obtained by running the model. 
Each line connects a set of data points corresponding to different $k$ values 
and represents the equilibrium spin periods at a given orbital period. The ten 
lines correspond to orbital periods of 80~m, 2~h, 3~h ... 10~h. In general the 
uncertainty in each equilibrium spin period is smaller than the size of the
plotted point, except for a few of the points at the lower extreme of the 
lines, where indicative error bars are shown. The measured magnetic moments
of polars are shown by the triangles along the top of the figure. The estimated 
magnetic moments of the intermediate polars are shown by squares and have been 
obtained by tracing across at the appropriate $P_{\rm spin}/P_{\rm orb}$ 
value to intersect the equilibrium line for the appropriate orbital period.}
\end{figure}

\clearpage

\begin{figure}
\caption{Examples of the accretion flows at equilibrium for $P_{\rm orb} =
4$~hr. The top row shows side-on views of the system, the bottom row shows the
same system in a face-on view. The left panels have $k=500$ and correspond to 
$\pspin / \porb = 0.07$
and $\mu_1 \sim 7 \times 10^{32}$~G~cm$^{3}$; the middle panels have $k=50000$ 
and correspond to $\pspin / \porb = 0.24$ and $\mu_1 \sim 7 \times 
10^{33}$~G~cm$^{3}$; whilst the right panels have $k=10^7$ and correspond to 
$\pspin / \porb = 0.53$ and $\mu_1 \sim 10^{35}$~G~cm$^{3}$. Truncated 
disc-like flows occur for lower magnetic moments at $\pspin / \porb < 0.1$; 
stream-like flows occur for higher magnetic moments at $0.1 < \pspin / 
\porb < 0.5$ and accretion fed from a ring-like structure at the outer edge 
of the WD Roche lobe occurs for the highest magnetic moments at $\pspin /
\porb > 0.5$.}
\end{figure}

\clearpage

\begin{figure}
\includegraphics[angle=-90,scale=0.6]{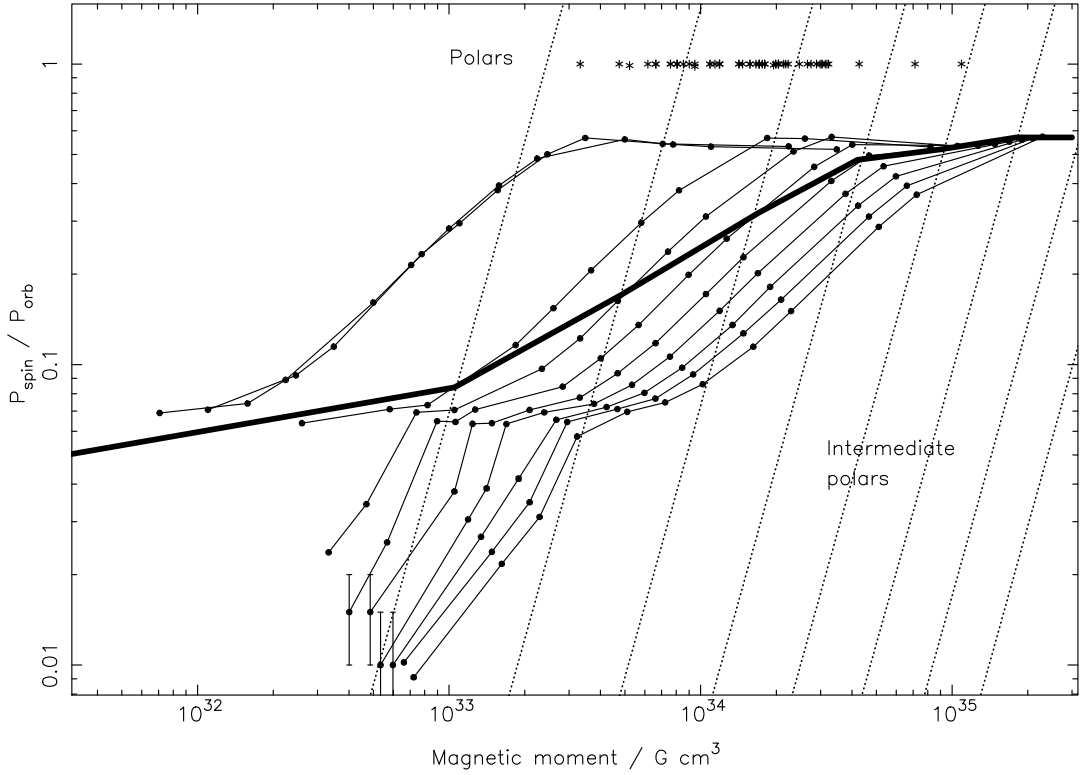}
\caption{Diagonal lines show solutions to Equation \ref{synch} for orbital 
periods 3~h, 4~h, ... 10~h. The lines for shorter orbital periods are to the 
upper left of the region plotted.  Where these diagonal lines intersect the 
equilibrium spin period lines for their respective orbital periods, gives 
the points at which systems will synchronize. The locus of these points is 
shown by the thick line which simply connects the various intersection points.}
\end{figure}

\clearpage

\begin{figure}
\includegraphics[angle=-90,scale=0.6]{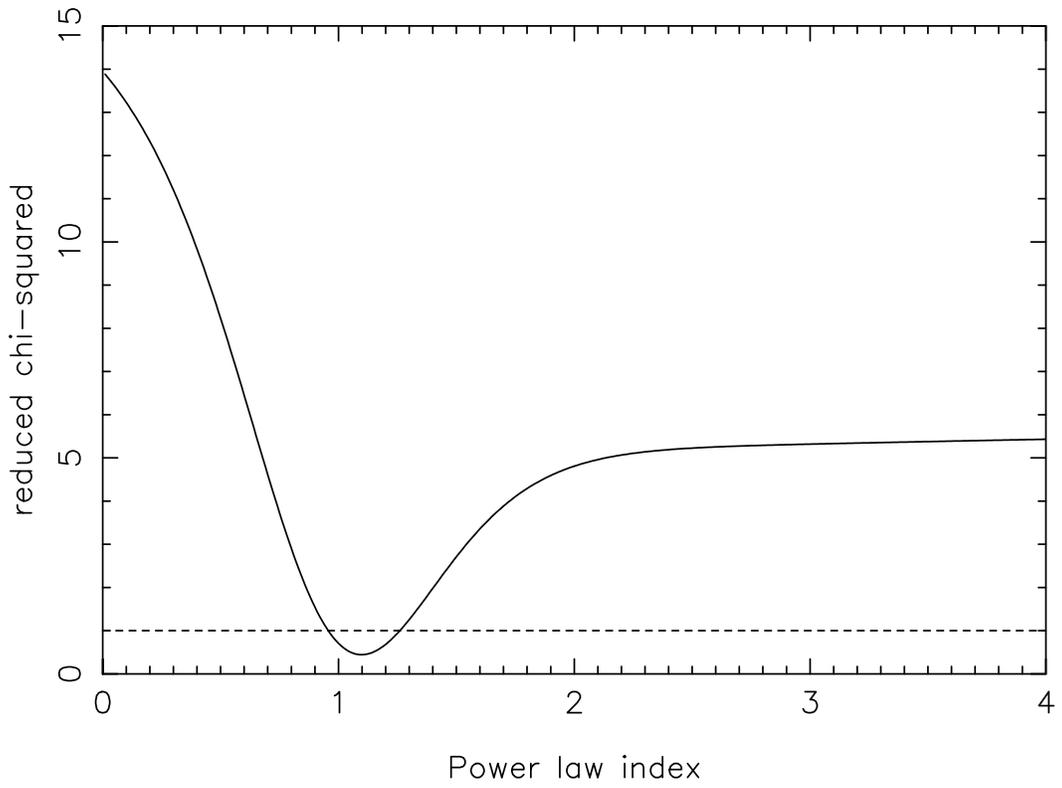}
\caption{The reduced chi-squared verses power law index obtained by fitting
Equation \ref{mu1} to the observed number of mCVs with $P_{\rm orb}>3$~h.}
\end{figure}

\clearpage

\begin{figure}
\includegraphics[angle=-90,scale=0.6]{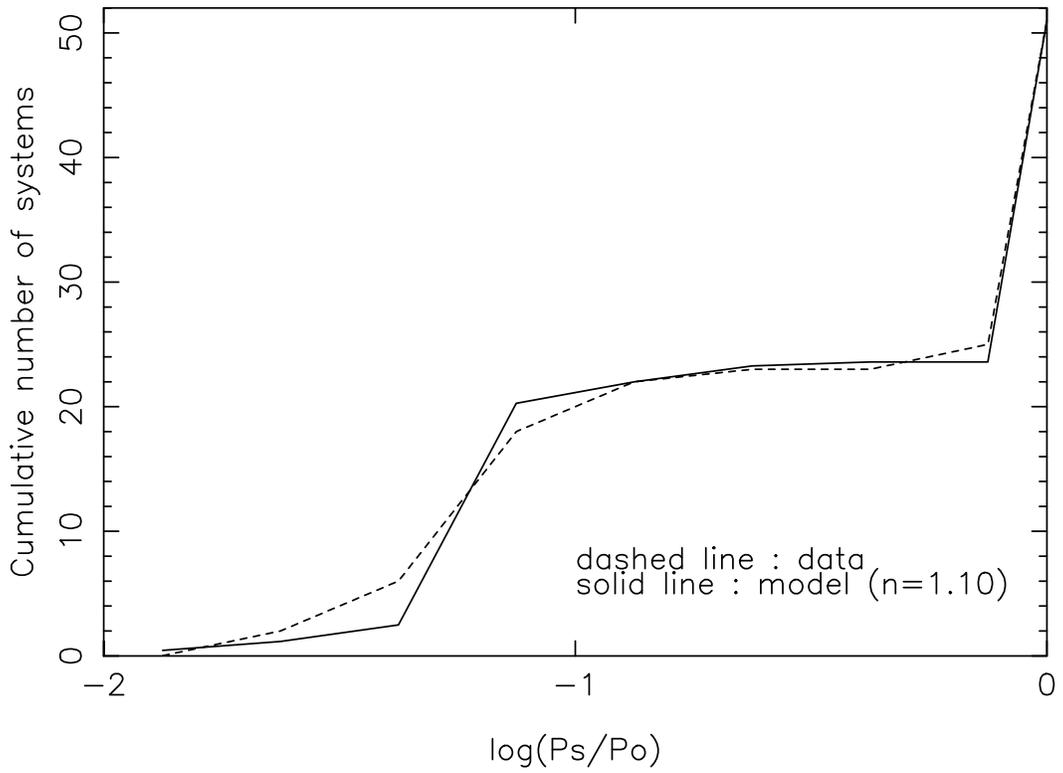}
\caption{Cumulative histograms showing the observed number of mCVs with 
$P_{\rm orb}>3$~h and the predicted number according to Equation \ref{mu1} with 
$n=1.10$, as a function of spin to orbital period ratio.}
\end{figure}

\end{document}